# Combining cortical voltage imaging and hippocampal electrophysiology for investigating global, multi-time scale activity interactions in the brain

## Authors

Rafael Pedrosa[1]* and Francesco Battaglia[1]

## Author Affiliation

[1] Donders Institute for Brain Cognition and Behaviour, Radboud University, Nijmegen, the Netherlands.

*** Correspondence:**

Rafael Pedrosa, Ms.C.

Radboud University

Donders Institute for Brain, Cognition and Behaviour

Neuroinformatica

Heyendaalseweg 135

6525 AJ NIJMEGEN

Internal postal code: 66

Email: rhapedrosa@gmail.com

## Acknowledgements

We thank Jeroen Bos for technical help, Arne Meyer for help with the visual stimulation experiments and Thomas Knopfel for donating the CaMK2A-tTA;tetO-chiVSFP mouse line. This work was funded by the European Union's Horizon 2020 research and innovation program (MGate, grant agreement no. 765549; F.P.B.), and the European Research Council (ERC) Advanced Grant ``REPLAY-DMN'' (grant agreement no. 833964; F.P.B.).

## Number of pages:

20

## Abstract

A new generation of molecular tools for analyzing neural activity has been contributing to elucidate classical open questions in neuroscience. With enhanced GEVIs and advanced optical imaging techniques, voltage-imaging technologies have been increasingly used to observe the dynamics of large circuits. In this chapter, we describe how to combine cortical wide-field voltage imaging with hippocampal electrophysiology in awake behaving mice. Furthermore, we highlight how this method can be useful for different possible investigations.

## 1 Introduction:

Hippocampo-cortical interactions have been widely studied to understand their role in supporting memory[1,2]. Yet, most studies have focused on the interplay between the hippocampus and a few cortical areas with major anatomical links to the hippocampus, such us the entorhinal or the medial prefrontal cortex[1,3–5]. Less is known about the connection between hippocampal activity on one hand, and activity phenomena spanning the cortex as a whole. Those global phenomena are typically studied with imaging techniques measuring indirect correlates of neural activity, such as the blood-oxygen-level-dependent BOLD or calcium intake, whereas most of our knowledge about the hippocampus come from the more direct view provided by electrophysiology. Cortical wide-field voltage sensitive (VS) imaging in Genetically encoded voltage indicator (or GEVI) mice instead provides a global view on cortical activity at the same time scale and measuring the same variable, membrane potential as electrophysiology. We describe here the combination of cortical VS imaging with hippocampal electrophysiology. In this case, multiple cortical areas can be recorded simultaneously, as well as the hippocampus, in a proportionate timescale.

### 1.1 Electrophysiological investigation of the Hippocampus observed with electrophysiology

Since the HM patient case was reported to the scientific community[6], and lately the discovery of place cells by John O`Keefe in 1971 was reported[7], the hippocampus became one of the most studied areas in the brain. Research on hippocampal activity has been carried out by using electrophysiology techniques[1,2,4,7]. Electrophysiology was the main mean of investigation for key hippocampal activity phenomena such as Sharp-wave ripple and gamma[8–10]. During some mnemonic processing, such events are likely to coordinate interactions between brain regions, promoting memory consolidation process[9,11,12]. Because of this, the link between hippocampal activity and activity in neocortex is an important subject of study.

### 1.2 Cortex observed with electrophysiology and neuroimaging

Global cortical activity cortex has been studied with transcranial electrophysiology (EEG) and functional imaging (fMRI). These commonly utilized noninvasive approaches for studying human brain activity, however, cannot offer the resolution needed to comprehend the mechanistic underpinnings of information processing. In particular, EEG lacks the spatial resolution to identify basic neuronal sources, it has the essential temporal resolution to examine the dynamics of brain function. Thus, Event-related potentials (ERP) and oscillatory responses are commonly studied in association with behavioral responses[13–16]. In contrast, fMRI temporal resolution is much slower than the time scale of neural processes supporting most perceptual and cognitive activities, but produces highly localized measurements of brain activation ( ~2 mm resolution), it has been used to explore complex functional and anatomical networks and their role in cognitive processes [17–19].

### 1.3 Wide-field voltage imaging of cortical neural activity

With the limitation of obtaining both spatial and temporal resolutions in the same data, new neurotechnologies have focused on this improvement. On the molecular approach, genetically encoded voltage indicators (GEVI) lines of mice in combination with wide-field optical imaging has shown to be a good spatiotemporal alternative to study cortical communication[20–22]. GEVIs have been used extensively to study the dynamical interplay between sensory and associative cortices[23,24]. Making use of it, classical studies were lately reproduced[2,22,25–27]. Voltage optical imaging offers significant advantages over classical approaches: (1) Temporal resolution comparable with electrophysiology. (2) Selective investigation of specific cell population of interest (for example: layer specificity and specific subclasses of interneurons and principal cells) and (3) Multiple imaging sessions in awake mice can be conducted over a period of days or weeks. Furthermore, this approach has boosted the study of cortical circuit dynamics in the framework of complex self-organizing systems[22,24,28,29].

## 2 Materials:

### 2.1 Animals

We used CaMK2A-tTA;tetO-chiVSFP transgenic mice (3 to 6 months and weighting 25-35g)[20,21]. This line of animal expresses a GEVI specifically in pyramidal neurons across all the cortical layers. For the macroscopic epifluorescence imaging, the optical electrophysiology signal was restricted to the superficial cortex[20,21,24,25]. The animals had ad libitum water and food access under a 12 hours light/dark cycle. All the experiments were approved and conducted in accord with the Dutch governmental Central Commissie Dierproeven (CCD) and European Directive 2010/63/EU on animal research.

## 3 Methods:

### 3.1 New technical approach for study mechanisms in neuroscience

To investigate the interactions between hippocampus and cortex we used GEVI line of mice to measure the instantaneous voltage activity of the entire neocortex profile in combination with a silicon probe in the CA1 hippocampus[2,25].

### 3.2 Procedure for combining wide-field imaging with hippocampal electrophysiology

To record the cortical optical imaging in combination with hippocampal electrophysiology, we performed a surgical procedure to expose the scalp and chronically implant a 16 channels linear probe (Atlas Neuroengineering, Belgium, 50 µm spacing between recording sites) in the hippocampus. Initially, the animals were placed in a stereotaxic with a nasal mask delivering isoflurane at 0.5-1.5%. During the entire surgery, the body temperature was kept by a heat-pad under his body around 37 °C and the breathing rate at 0.5-1 Hz. Then, lidocaine at 2% (50 µl) was injected subcutaneous at the incision zone. Successively, we exposed a wide area of the scalp (Fig. 1a), and we thinned the skull of the entire right hemisphere with a surgical drill in order to remove the capillaries, and reduce light scattering due to bone tissue. Additionally, 2 screws were implanted in the left hemisphere for a better fixation of the head-plate, together, a third screw was also implanted in the superficial region of the cerebellum in the same hemisphere to be used as a ground and reference. For hippocampal electrophysiology recording, a 16 channels linear probe (Atlas Neuroengineering, Belgium, 50 µm spacing between recording sites) was chronically implanted in the right hemisphere (ipsilateral from the imaged cortex) (Fig. 1b). The probe was initially placed in +2.4 mm ML and -4.3 mm AP at a 60 degree angle, which was then lowered and pinned at ~2.2 mm depth. Finally, the head-plate was carefully placed and fixed on the scalp with acrylic cement.

After the animal recovered from the surgery, we imaged the voltage activity from the neocortex. To measure the both mKate2 (GEVI FRET acceptor) and mCitrine (GEVI FRET donor) epifluorescence, we used a macroscope similar as described by Song et al., 2018 (Fig 1c-d). The wide-field images were acquired by two synchronized sCMOS PCO edge 4.2 cameras in combination with Leica PlanAPO1.0 and Leica PlanAPO1.6 lenses. The fluorescence excitation light was provided by high power halogen lamps (Brain Vision, Moritex). The Semrock optics attached to the macroscope were: mCitrine excitation 500/24 and emission FF01-542/27, mKate2 emission BLP01-594R-25 with beam splitters 515LP and 580LP. To compute the voltage activity, a ratio ($\Delta R/R$) between the mCitrine–mKate2 was calculated ($R$ = mCitrine/mKate2 signals), then normalized as $\Delta R/R = (R(t) - R_{mean})/R_{mean}$[20,25]. In parallel, the local field potential (LFP) from the hippocampus was acquired using a 16- channels headstage (Intan Technologies, RHD2132) and an Open Ephys system. To record, the raw signal was filtered in 0.1-7500 Hz and sampled in 20kHz and downsampled to 1kHz for the analysis. For the synchronization between the cameras (2 for widefield imaging and 1 for eyetracking) and the electrophysiology data, we recorded pulses from the cameras, signaling time of acquisition of each from in an analog channel of the Open Ephys system.

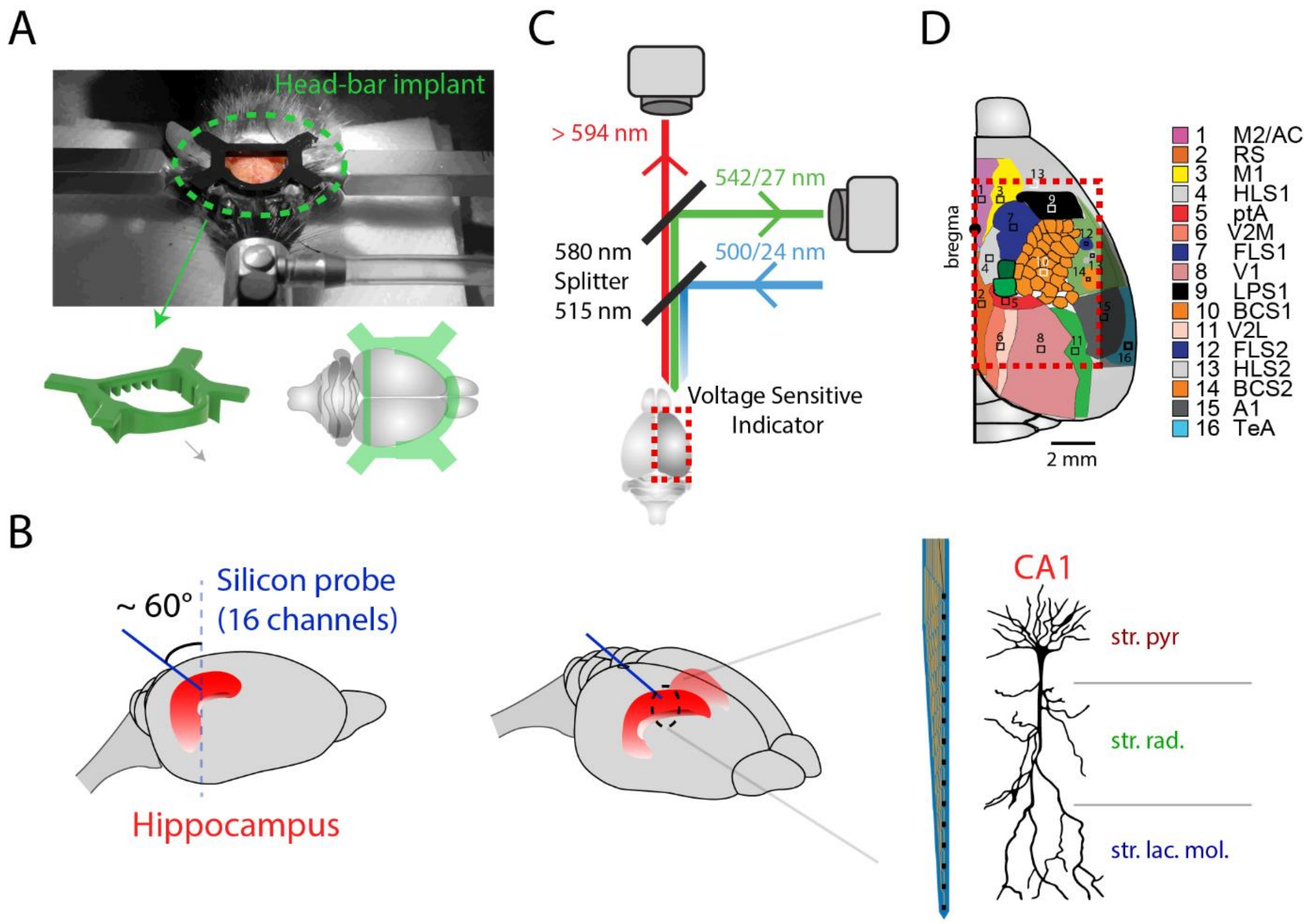

**Fig.1** Surgical procedures for the wide-field cortical imaging and hippocampal electrophysiology recordings. **(a)** Head-bar implantation on a mouse placed in the stereotaxic apparatus. **(b)** Schema of a high density silicon probe (16 channels) implantation in the hippocampus. With a 50 µm spacing between the contacts, the probe can cover all the layers in CA1. Note that the implant is done in the posterior part of the cortex, specifically close to the lambdoid suture, thus leaving the entire cortical area free to be imaged. **(c)** Ratiometric mesoscope used for neocortical optical imaging. The fluorescence emitted from the superficial cortical layers were recorded by two cameras. The ratiometric measurement of the cameras result in the voltage variation of the recorded area. **(d)** Topographic map of the multiple areas that can be imaged together by the mesoscope (topographic image by Majid Mohajerani).

### 3.3 Cortico-hippocampal signal modulation during behavior

To visualize the cortical voltage activity with the hippocampal LFP, we recorded two vital states, ex vivo (dead animal) and in vivo (alive in resting state) from the same mouse in order to compare the signal-to-noise responses. For this, we acquired images from the right neocortex at 50 Hz at 375 ×213 pixel (~60 pixel/mm) and 12 bit sampling depth in blocks of 10 minutes. As expected, in the ex vivo condition, the neocortex does not show any activity, whereas in the living brain spontaneous fluctuations are manifest in the neocortex (Fig. 2a). Simultaneously, spontaneous Sharp-wave ripple (SWR) events can be observed in the hippocampal pyramidal layer around the cortical transients as shown in the example in figure. As already reported, the spontaneous emergence of hippocampal SWR are probabilistically linked to specific cortical areas[2,30], consistent with our data.

The ratiometric approach allows us to control certain types of physiological influences, such as hemodynamic influences and heartbeats, the outstanding fluorescence corresponds to the domain of the signal-to-noise. For this correction, we used a gain equalization procedure in both mKate2 and mCitrine fluorescence before the ratio calculation in order to isolate heart beat components in addition to a high pass filter <0.5 Hz to remove hemodynamic signals [21,25]. The resulting signal can be used to get a precise picture of the cortical dynamical state. As an example, we report here the frequency distribution of the size of neocortical transients, large fluctuations involving a significant portion of the imaged cortical territory. For this, we detected the transients by using a similar approach reported by Scott et al., 2014 [31], where the global state represented by the active states of each pixel has to cross a certain threshold defined as the mean of the total activated frames. As the result shows, the in vivo distribution shows distinctive features (Fig. 2b). In particular, a higher proportion of large transients can be observed in the in vivo animal compared to ex vivo.

The impact of movements in the active behavioral state causes a strong modulation in the cortex-wide neural activity[32]. This important finding has been strongly considered in recent studies[33–36]. In the hippocampus, theta frequency band (5-10 Hz) is a hallmark of active states, such as walking, running and sniffing. Consistent with previous studies[32,36] we found that cortical activity in multiple brain areas increases during bouts of treadmill running. Interestingly, coexisting with hippocampal theta, cortical activity, mostly primary motor cortex (M1), seem more tightly related to running (Fig. 2c).

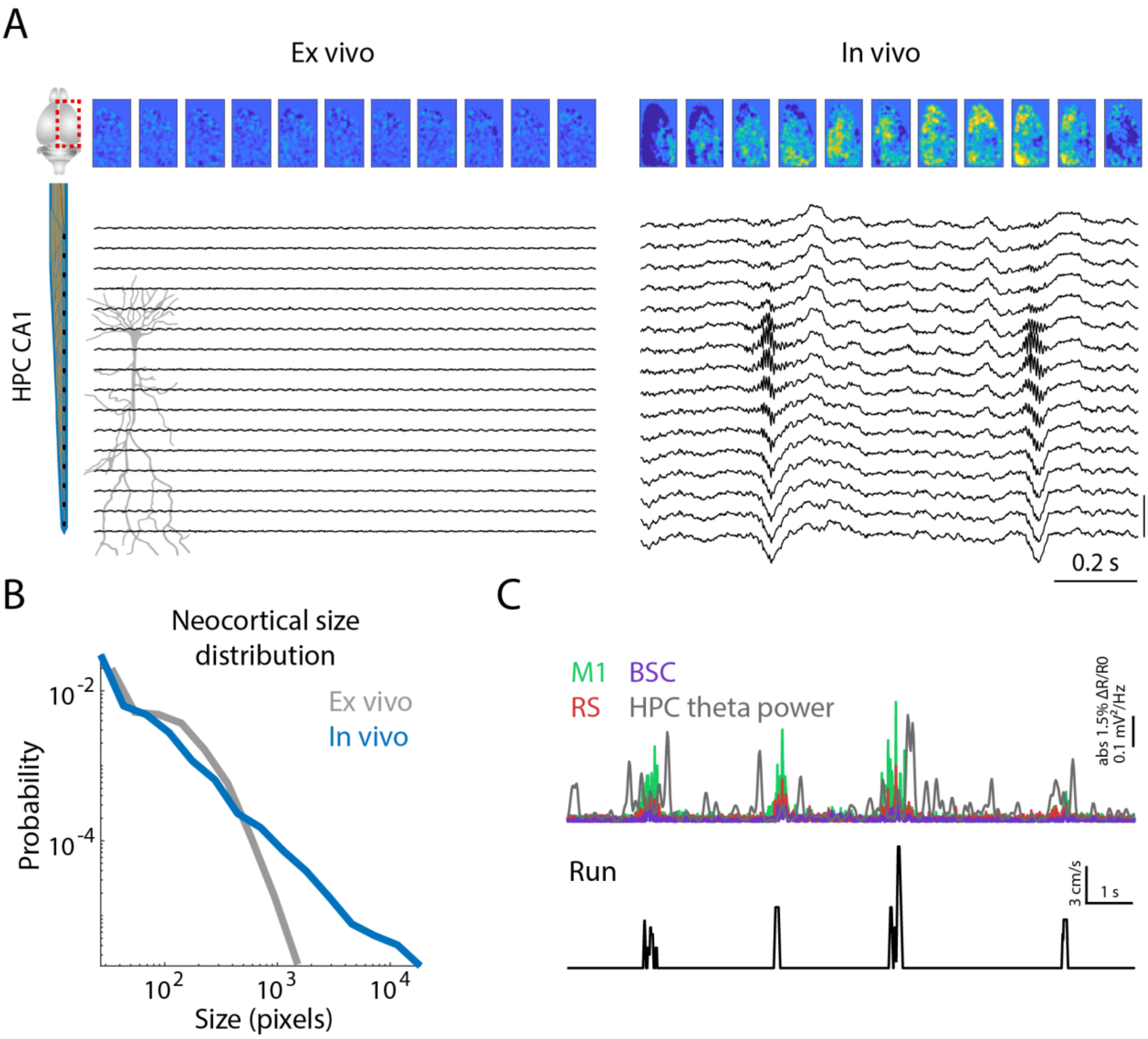

**Fig.2** Signal quality and vital state modulation of cortical activity. **(a)** Examples of neocortical imaging and hippocampal CA1 electrophysiology during ex vivo and in vivo states of the animal. **(b)** Probabilistic distribution of neocortical transients based on the sizes (the spatial extent of the activity fluctuations) for ex vivo and in vivo states. As a comparative proportion, and expected, the in vivo state obeys a distribution more likely to a power-law scale. Smaller transients, also present in the ex-vivo preparation, are likely due to noise. **(c)** Comparative absolute ΔR/R cortical activity with hippocampal theta power during quiet and active behaviors. Note that during the run periods, the animal presents a higher modulation of the signal, which is comparative with the increasing of theta power in the hippocampus.

## 3.4 Layer-specific hippocampal signal during integrative visual information process

Visual information plays an important role in spatial navigation [37,38]. Particularly, identifying landmarks and borders are processes that require information from primary visual areas, to reach higher hierarchy areas such as the hippocampus in the brain where a cognitive representation of the environment may be generated. Nevertheless, how visual signals modulate spatial representations and how information propagates to the hippocampus is still unknown. To investigate this, we advocate that wide-field recordings of the cortex, may provide important details on the visual integration mechanism. Towards this goal, we show here that the activity from visual cortex propagates to other higher cortical areas that have direct interconnections, such as retrosplenial, cingulate and parietal cortices (Fig. 3a-b)[22,39].

In rodents, eye movements have also been linked to spatial navigation processes[40–43]. This is quite relevant, because instantaneous visual input constantly updates egocentric and allocentric spatial information and gaze direction is a vital input to that computation. Combining eye tracking with hippocampal probe recordings allows the investigation of layer-specific processes regarding the visual perception of the animal (example of the saccade movements, pupil size and the location the animal is focusing). Using a fixed camera (Basler aca1920-150um, lens of 20mm) attached to an infrared sensor, we could track pupil position with DeepLabCut[44], and we could detect the saccade movements of the eye. We observed correlated activity in the hippocampal probe, with a certain modulation on the spectral component (Fig. 3c). Interestingly, we specifically observed an emergence of medium gamma after the saccade movement on the channel located at *stratum lacunosum moleculare*. Besides that, this modulation seems also to be different across layers, which suggests that the saccade movement of the eye influence specific local processes within the hippocampus.

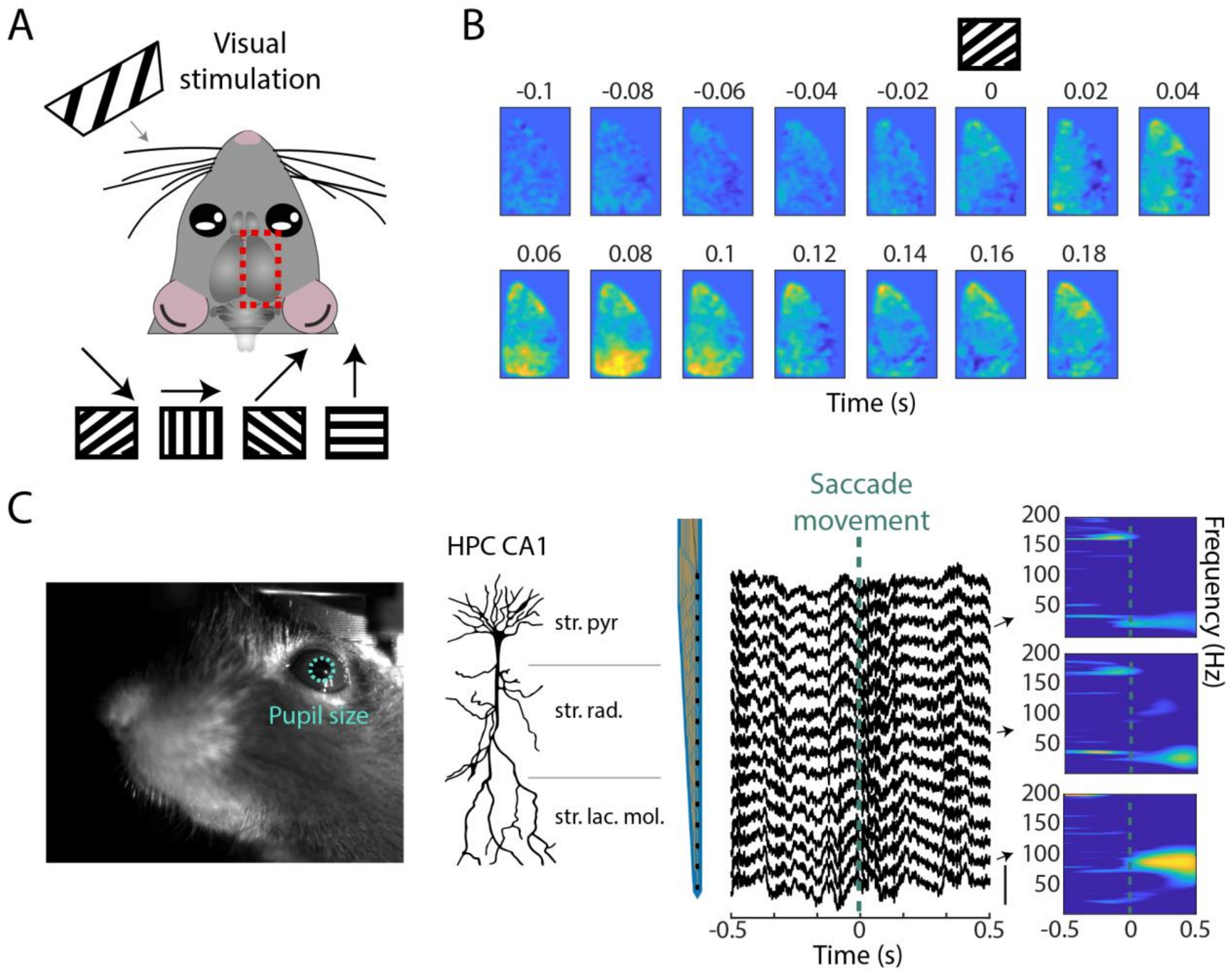

**Fig.3** Visual information propagates to cortical association areas and the hippocampus. **(a)** Schema of the visual stimuli paradigm used for this experiment. **(b)** Averaged voltage profile from the neocortex during specific visual stimulation pattern (45°). **(c)** Facial camera with the pupil size detected by DeepLabCut (left panel). On the right panel, the average local field potential of the hippocampal silicon probe around the saccade movement of the eyes. The spectrogram shows the spectral average of 3 channels (each located in *stratum pyramidale, stratum radiatum and stratum lacunosum moleculare*) around the saccade movement of the eye.

## 3.5 Hippocampal neuronal interrelation with cortical modules

Hippocampus is seen as an important central area for the memory process. Therefore, it is composed of several efferent and afferent projections scattered throughout the brain, among them the cortex. The classic hypothesis that the hippocampus retains short-term memories and indexing them to the cortex for long-term memory, highlights the importance of hippocampo-cortical communication[1,45]. Prior to this process, a short-term memory emerges from the encoding of sensory information and its adaptation to the hippocampus. In this sense, the spike modulation resulting from the hippocampo-cortical interaction has room for to be explored. In this section, we combined the cortical voltage activity with hippocampal units in order to see how this modulation could occur. Initially, we detected and spike sorted the hippocampal probe data by using kilosort 2.0 [46](Fig. 4a). Multiple neurons could be detected from the same recording sites, with distinctive waveforms (Fig. 4b). Next, we computed spike-triggered averages of the neocortical activity referred to hippocampal spiking activity hippocampus (Fig. 4c). In the examples below, we can see 3 neurons that correlate to different cortical areas (examples: 1-visual cortex, 4 parietal and retrosplenial cortices and 6 medium line). This shows the potential of this method to investigate the cortico-hippocampal interaction, in particular when firing of hippocampal place cells and its cortical correlates are taken into account. Probes with higher site density (eg Neuropixels[47]) may probe very useful for this.

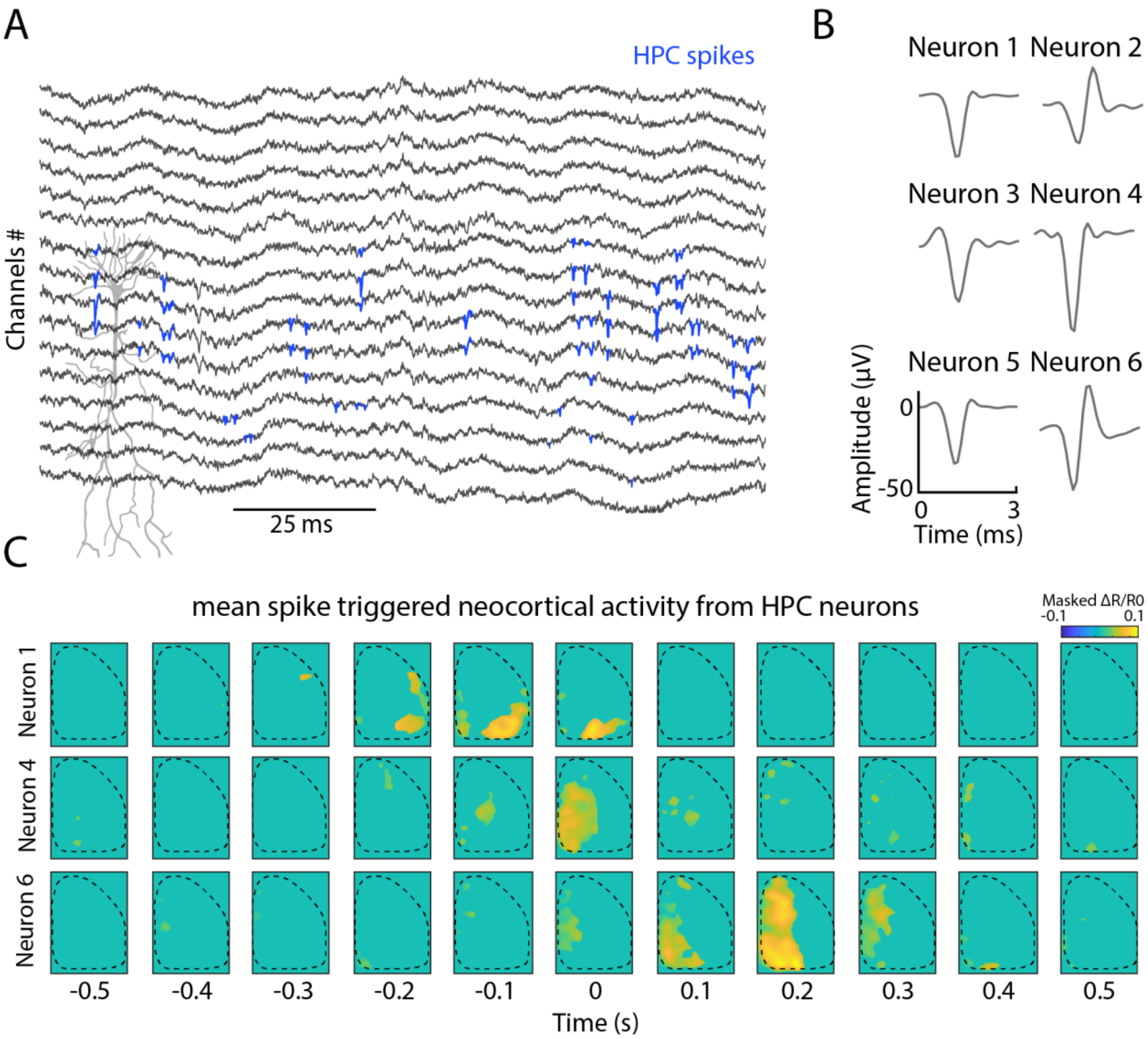

**Fig.4** Hippocampal neurons interact with specific cortical structures. **(a)**. Local field potential from the silicon probe in the hippocampus. In blue, putative spikes detected by kilosort 2.0. **(b)** Example of the waveforms from 6 neurons detected in the pyramidal layer. **(c)** Neocortical activation around spike activity from 3 neurons in the hippocampus. The plots are masked by the mean of the signal plus 5 standard deviation.

## 3.6 Cortical microstimulation and hippocampal electrophysiology

For almost a century, electrical stimulation has been utilized to better understand the natural circuitry of the brain[48,49]. It has been also used to look at a variety of cortical functions and connections, including hippocampal networks[50,51]. In this section we used the combination of wide-field voltage imaging with hippocampal electrophysiology to investigate how local cortical microstimulation affects hippocampal oscillations. More precisely, we electrically stimulated Retrosplenial (RS) and Somato-Sensory (SS) cortices with distinct current intensities (20, 50, 100 and 200 μA) to observe how these two structures interact with hippocampal oscillations at different frequencies in a resting state mouse (Fig. 5a).

For the cortical microstimulation, we used a bundle of two tungsten electrodes (50 μm diameter) that were staggered to create a vertical distance of ~250 μm between the tips. The surgical procedure had the same steps as the one describes on the 3.2 section with the addition of a chronic implantation of two stimulation electrode bundles in RS and SS before the head-bar fixation. For RS implantation, we first performed a small craniotomy on the right hemisphere skull about +0.5 mm (ML) and -2.5 mm (AP). Then, we implanted the bundle electrode in +0.5 mm (DV) and 45 degrees towards the left hemisphere (outside of the visual field of the camera for the cortical imaging). For the following step, we again performed a small craniotomy on the right hemisphere at +3 mm (ML) and +1 mm (AP) above the SS cortex. The bundle electrode in this case was also implanted +0.5 mm (DV) but angle in 30 degrees towards the lateral side of the brain.

After 10 days, when the animal recovered from the surgery, we placed it in the setup for the cortical stimulation. For that, we used an Arduino UNO to send 50 square pulses of 10 milliseconds with inter-stimulus interval randomized between 1.5-2.5 seconds controlling a stimulus Isolator (World Precision A365), which was also connected to the Open Ephys for the synchronization. Observing the averaged wide-field cortical imaging around the RS and SS stimulations (at 200 μA), we found in the stimulated area an instantaneous burst of activity followed by ‘spreading depression’ lasting for about one second (Fig. 5b). Looking at stimulation effects on the hippocampal spectral activity on different intensities (20, 50, 100 and 200 μA), we found that RS stimulation (suppression) caused an interference on the hippocampal LFP (mostly in frequencies between 5 and 40 Hz) (Fig. 5c). This hippocampal oscillatory interference is proportional and statistically significant with the size of the current (and consequently RS suppression) induced in the RS area (Fig. 5d). Conversely, SS stimulation did not show significant difference in the hippocampal LFP (repeated-measures ANOVA, n = 50 stimuli for each condition). Combining cortical imaging, cortical microstimulations and hippocampal LFP, our result suggests that a causal relationship may exists between RS and hippocampus, but for that a further investigation has to be done, for example to establish if this is due to cortical inputs into the hippocampus, or to antidromic activation and suppression of hippocampal afferents in retrosplenial cortex.

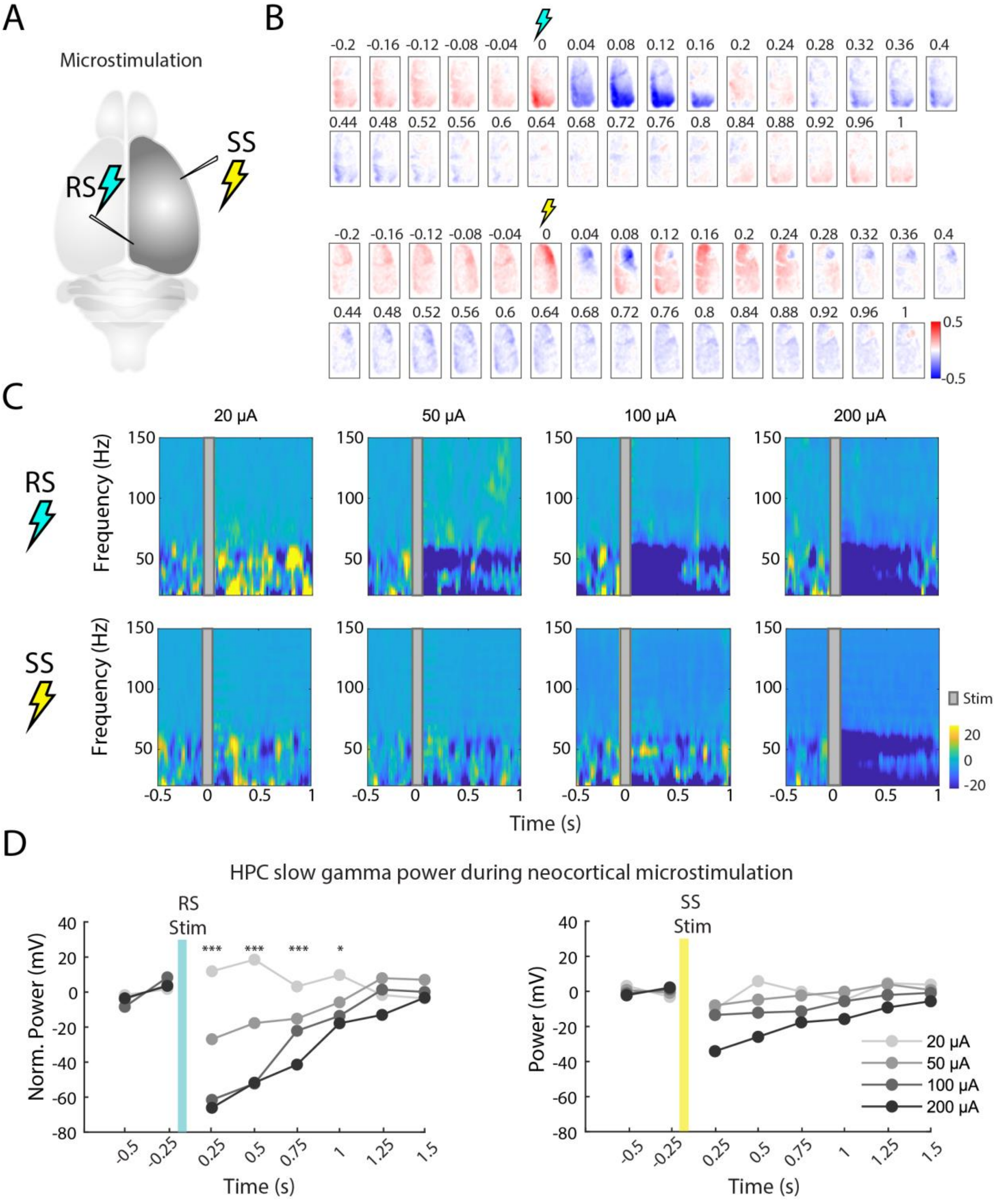

**Fig.5** Artificial inhibition in Retrosplenial cortex influences Hippocampal gamma power. (A) Schematic of the microstimulation zone for the RS and SS cortices. (B) Triggered wide-field VSI signal for RS and SS stimulations (10 ms stimulation; 200 µA; average of 50 stimulation events). Note that after the short period of positive activity, a continuous suppression takes place along the subsequent second. (C) Hippocampal CA1 spectrogram during 20, 50, 100 and 200 µA current for the RS and SS stimulation. (D) Baseline normalized power of gamma (20-50 Hz) around the 20, 50, 100 and 200 µA stimulations (*$p<0.05$, and ***$p<0.0005$, repeated-measures ANOVA, n = 50 stimuli for each condition).

## 4 Conclusions:

The study of complex brain processing into behavioral responses requires a method that monitors large-scale activity with high temporal precision. Using a GEVI mouse line, we combined wide-field voltage imaging of the neocortex with a silicon probe in the hippocampus CA1 while the animal rested or performed traditional behavioral tasks. This opens a new horizon to understand how sensory integration relates to the hippocampal activity at a behavioral timescale.

A common mistake committed in neuroscience is the assumption of ascribing specific cognitive responses only to certain brain areas. This is due to the fact that animal research is commonly designed to target specific target areas, leaving the contributions of the rest of the brain aside. Here, we advocate that wide-field voltage imaging can be a valuable instrument to observe multiple cortical areas simultaneously in awake behaving animals. Compared to other molecular indicators, such as: calcium and glutamate, voltage activity has a faster temporal resolution, similar to that of electrophysiology, which allows causal analysis with LFP data.

It is important to emphasize that the combination of electrophysiology with cortical voltage imaging is flexible and may be adapted to different scientific questions. An example could be asking how neuronal ensembles in a given brain area correlate with broad neocortical networks. In this case, a neuropixel[44] instead would be a better fit and could be easily replaced by applying the same methodological procedure reported here.

Finally, we believe that for the next few years, cortical wide-field imaging will contribute to the understanding of hierarchical neocortical processes in different behavioral states.

Combined with hippocampal LFP, this approach has the potential to explain fundamental circuitry of cognitive processes such as memory consolidation.